\documentclass[conference]{IEEEtran}
\usepackage{blindtext, graphicx}
\usepackage{listings}
\usepackage[keeplastbox]{flushend}
\ifCLASSINFOpdf
\else
\fi
\usepackage[font=footnotesize]{subfig}

\usepackage{color}

\usepackage{fancyhdr}

\begin{document}
%
\title{Enhanced Socket API for MPTCP \\ \LARGE{Controlling Sub-flow Priority}}


\author{\IEEEauthorblockN{Abhijit Mondal}
\IEEEauthorblockA{Department of Computer Science \& Engineering\\
Indian Institute of Technology Kharagpur\\
Kharagpur, India 721302\\
Email: abhimondal@iitkgp.ac.in}
\and
\IEEEauthorblockN{Aniruddh K}
\IEEEauthorblockA{TCS Research \& Innovation\\Bangalore\\
India 560066 \\
Email: aniruddh.k@tcs.com}
\and
\IEEEauthorblockN{Samar Shailendra}
\IEEEauthorblockA{TCS Research \& Innovation\\Bangalore\\
India 560066 \\
Email: s.samar@tcs.com}
}



\maketitle
\thispagestyle{fancy}

\begin{abstract}
Multipath TCP (MPTCP) can exploit multiple available interfaces at the end devices by establishing 
concurrent multiple connections between source and destination. MPTCP is a drop-in replacement for 
TCP and this makes it an attractive choice for various applications. In recent times, 
MPTCP is finding its way into newer devices such as robots and Unmanned Aerial Vehicles 
(UAVs). However, its usability is often restricted due to unavailability of suitable socket 
APIs to control its behaviour at the application layer. In this paper, we have introduced several 
socket APIs to control the sub-flow properties of MPTCP at the application layer. We have proposed 
a modification in MPTCP kernel data-structure to make the sub-flow priority persistent across 
sub-flow failures. We have also presented Primary Path only Scheduler ($PPoS$), a novel sub-flow 
scheduler, for UAVs and similar applications/devices where it is necessary to segregate data on 
different links based upon type of data or Quality of Service (QoS) requirements. We have also 
introduced the socket APIs for providing the fine grained control over the behaviour of $PPoS$ for 
particular application(s) rather than changing the behaviour system wide. The scheduler and the 
socket APIs are extensively tested in Mininet based emulation environment as well as on real 
Raspberry Pi based testbed. 
\end{abstract}

\section{Introduction}
\label{intro}

Today, it is difficult to imagine life without the Internet even for a short span of time. Several 
research proposals have been made over time to improve the reliability of the Internet. Multipath 
TCP (MPTCP) \cite{rfc6182,rfc6824} is one of the such attempts to improve the throughput and 
reliability by leveraging multiple physical interfaces available on the end device. With standard 
TCP/IP networking stack, a device can not utilize multiple network interfaces simultaneously, as 
TCP is tightly coupled with the underlying network layer and it can communicate between a pair of 
network addresses only. In contrast with TCP, MPTCP can utilize multiple network interfaces to 
communicate with a remote host. The concurrent use multiple paths improves the resilience to the 
network failure by continuing the 
transmission of data on other available paths\footnote{In this paper, MPTCP paths and MPTCP 
sub-flows refer to the same thing (represented by a combination of source IP, source port, 
destination IP and destination port) unless specified otherwise explicitly.}. It can also provide 
seamless support for mobility by allowing run time change in the network address during the 
handover process \cite{boucadair2016mptcp}.

MPTCP is demonstrated to be useful primarily in datacenters for large data transfers 
\cite{Raiciu2011IDP}. However, MPTCP can be quite helpful in day-to-day communication as well. All 
modern smart-phones contain at least 
two network interfaces i.e. cellular and WiFi. These devices can use MPTCP to transmit data 
simultaneously using both the interfaces 
\cite{chen2013measurement,Paasch2012Sigcomm,Lim2014AllThingsCellular,Raiciu2012NSDI}. This has  
motivated many industries to include MPTCP in Android/IOS based smart-phones \cite{seo2015giga} so 
that it can use both the interfaces to download data at higher speed with improved reliability. 
Although most of the legacy servers do not have MPTCP deployed, there is SOCKS proxy 
available which can help accessing any server using MPTCP \cite{rfc1928}. MPTCP has also been 
demonstrated, in the literature, to be used to improve network reliability in vehicular 
communication \cite{Williams_multipathtcp}.

Unmanned Aerial Vehicles (UAV) or drones (like a quad-copter, hexa-copter, octa-copter etc.) are 
gaining popularity these days. These devices are being used for various activities such as 
surveillance, security, rescue operations etc.. Most of the times, a UAV is connected to an 
access point using WiFi via standard TCP/IP link. The access point controls the UAV using this 
link with dedicated TCP connection while the UAV/drone sends a feed of live sensor data (like 
live video) to its access point. These control messages are sent from the access point to the UAV. 
Unlike the live data feed, these control messages are crucial to the proper behaviour of the UAV 
and should not be delayed or dropped. Moreover, the live sensor feed sends a huge amount of data 
over the physical link and causing the control messages to be delayed. Any delay in these control 
messages can cause severe damage to the UAV itself and jeopardize the entire mission.

One of the possible remedies to the above problem is to use multiple TCP/IP links between a base 
station and the UAV. One of the links can be dedicated for control messages while other for 
live sensor feed. This setup may work perfectly when both the links are alive. However, while a 
UAV is roaming around, it may happen that one link has failed or become unreachable momentarily. 
In such a scenario MPTCP, can easily move its connection from one interface to another transparent 
to the user and the underlying application.
 
While MPTCP can use all the available network interfaces concurrently to improve the throughput and 
reliability, it still does not solve the problem of delay in delivery of control messages to the 
UAV/drone 
because MPTCP uses and may congest all the available links simultaneously. To mitigate this 
problem without loosing its salient feature of improved reliability, we proposed 
\cite{Rao2017MPTCP,samar2016mptcp} a modification in MPTCP scheduler. According to the proposed 
modification, an application can mark one of the sub-flows as primary one and other sub-flows as 
backup. Only the selected primary sub-flow(s) is used for data transmission while in the event 
of failure of the primary one, MPTCP seamlessly starts using the alternate available sub-flows for 
the transmission. As soon as the former is restored, MPTCP reverts back to the same. 

In its current form, MPTCP has no socket API to provide this control to the applications e.g. Robot 
Operating System (ROS) in our case. In this paper, we have developed new socket APIs to provide a 
fine grained control at the application layer. Using these socket APIs, an application can 
dynamically modify the socket properties and prioritize one sub-flow over the other for its 
purpose. 
For example, in case of UAV/drones, ROS can prioritize one of the sub-flow for control messages, 
while 
other available ones for live feed. We have developed the socket API to enable/disable 
our proposed scheduler from the application layer on the need basis and also integrated the same 
with ROSTCP to expose these functionalities at the ROS layer.

Rest of the paper is organized as follows: In Section \ref{litsur}, we describe the background of 
MPTCP. In Section \ref{socketapi}, we discuss the details of new socket APIs. Section 
~\ref{drone} introduces the MPTCP kernel modifications for our proposed scheduler and the socket 
API changes for the UAV/drone scenarios. Section \ref{exp} provides experimental results with 
MPTCP. Finally, Section \ref{concl} concludes the paper.

\section{Background and Related Works}
\label{litsur}

\begin{figure}[!h]
	\centering
	\includegraphics[scale=0.45]{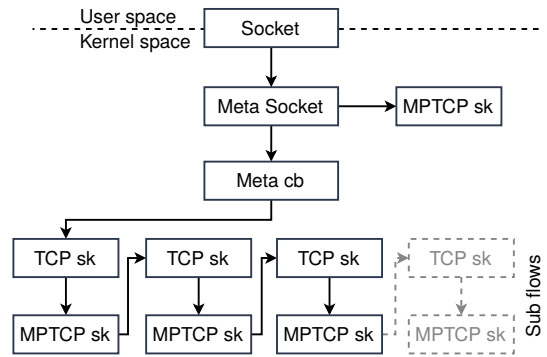}
	\caption{MPTCP socket framework}
	\label{fig:mptcp_framework}
\end{figure}

MultiPath TCP \cite{rfc6824} is a drop-in replacement for TCP. The existing applications can take 
advantage of MPTCP without any change in their implementations. MPTCP provides the same 
interface for connection initiation and communication between two hosts as that of TCP. When an 
application initiates a TCP socket in a MPTCP enabled system, it gets a standard TCP socket 
reference. Subsequently, this socket attempts to establish the connection to the remote host as an 
MPTCP connection. If the remote host is also MPTCP enabled, the reference socket turns into an 
MPTCP socket seamlessly transparent to the original application. We have depicted the MPTCP socket 
structure in Fig.~\ref{fig:mptcp_framework}. As shown in Fig.~\ref{fig:mptcp_framework}, MPTCP 
needs few more data structures than just the socket reference available to an application. The 
\texttt{Meta cb} which is a part of \texttt{Meta socket}, holds information related to MPTCP. This 
reference does not exist before MPTCP connection is established. The \texttt{Meta cb} structure 
contains a reference to all the sub-flows in the socket linked list. The \texttt{MPTCP sk} 
structure contains MPTCP related information for a single MPTCP sub-flow between source and 
destination. 

MPTCP implementation consists of several modules namely a) path-manager, b) scheduler and c) 
congestion control.

\subsubsection{Path-manager} Path-manager handles the connection between two end hosts. 
Currently, there are four path-managers being defined for MPTCP i) default, ii) full-mesh, iii) 
ndiffports, iv) binder. `default' path-manager does nothing more than accepting the passive 
creation of sub-flows.  `full-mesh' creates a full-mesh of sub-flows between all available source 
and destination interfaces. `ndiffports' creates multiple sub-flows between every source and 
destination pair while `binder' is based on Loose Source  
Routing\cite{Boccassi:2013:BSA:2502880.2502894}. It is interesting to note that in MPTCP, for an 
application, there can exist $m\times n$ possible combinations of interconnects where $m$ be the 
number of interfaces available at the source and $n$ at the destination. MPTCP can create multiple 
sub-flows on each interconnect by modifying the source port transparent to the application.

\subsubsection{Scheduler} MPTCP scheduler is responsible for scheduling packets among the active
sub-flows. There are multiple schedulers being defined for MPTCP. They schedule the packets based 
on different parameters such as Round Trip Time (RTT), Congestion Windows etc.. Note that if 
\texttt{low\_prio} MPTCP kernel flag is enabled for a sub-flow, it will be considered as a backup 
sub-flow and will be used to transfer data only if no active sub-flows exist.

\subsubsection{Congestion Control} Like standard TCP, MPTCP also has congestion control module. 
However, due to coexistence with TCP, MPTCP undergoes fairness issues \cite{rfc6356, 
OLIARamin2012}. Hence, several new congestion control algorithms are being proposed for Linux-based 
MPTCP implementation \cite{walid-mptcp-congestion-control-04, xu-mptcp-congestion-control-05}.

As per the current implementation, new sub-flows are created only in the client. Every newly created 
sub-flow is treated as active sub-flow and data is being transmitted over all the active sub-flows 
simultaneously. MPTCP have the framework to mark sub-flows as low priority sub-flows and send this 
information to the remote host with the help \texttt{MP\_PRIO} header option \cite{rfc6824}. 
Low-priority 
sub-flows are considered as backup sub-flows, and these are used to transmit data only when no 
active sub-flow is available. The limitation of the current MPTCP implementation is that such 
marking of a sub-flow 
is a system wide change and affects all applications running in that system. Moreover, most of 
the MPTCP related properties are also not accessible to the  
application. There are very few socket APIs available (Table~\ref{table:mptcp_socket_apis}). This 
table also includes the APIs recently published by Hesmans et. al. \cite{hesmans2016enhanced}.

\begin{table*}[t!]
\caption{\label{table:mptcp_socket_apis}List of avaible MPTCP socket APIs.}
\centering
 \begin{tabular}{|l|c|c|p{.5\textwidth}|} 
 \hline
 Name & Input & Output & Description \\ [0.5ex] 
 \hline\hline
 \texttt{MPTCP\_ENABLED} & true-false & - & Enable or disable the MPTCP from an application \\ 
\hline
 \texttt{MPTCP\_PATH\_MANAGER} & path manager & - & Set the path manager from an application \\ 
\hline
 \texttt{MPTCP\_SCHEDULER} & scheduler & - & Set scheduler for a MPTCP socket from an application 
\\ \hline
 \texttt{MPTCP\_INFO} & - & mptcp info & Get all MPTCP related information \\ \hline
 \texttt{MPTCP\_GET\_SUB\_IDS} & - & sub-flow list & Get the current list of sub-flows viewed by 
the kernel \\ \hline
 \texttt{MPTCP\_GET\_SUB\_TUPLE} & id & sub tuple & Get the pair of ips and ports used by the 
sub-flow identified by id \\ \hline
 \texttt{MPTCP\_OPEN\_SUB\_TUPLE} & tuple & - & Request a new sub-flow with pair of ip and ports \\ 
\hline
 \texttt{MPTCP\_CLOSE\_SUB\_ID} & id & - & Close the sub-flow identified by id \\ \hline
 \texttt{MPTCP\_SUB\_GETSOCKOPT} & id, sock opt & sock ret & Redirect the getsockopt given in 
input to the sub-flow identified by id and return the value returned by the operation \\ \hline
 \texttt{MPTCP\_SUB\_SETSOCKOPT} & id, sock opt & - & Redirect the setsockopt given in input to the 
sub-flow identified by id.\\
 \hline
 \end{tabular}
\end{table*}

From Table~\ref{table:mptcp_socket_apis} we can note that there are no APIs available to set or 
update priority or any other related properties of an MPTCP sub-flow. 

\section{Socket APIs for MultiPath TCP}
\label{socketapi}

In this section, we provide the details of socket APIs designed and developed to control the 
sub-flow priority behavior in MPTCP. These socket APIs have inherited sub-flow id related 
specifications from \cite{hesmans2016enhanced}. Using our socket API, we provide the control over 
scheduling of data transmission through different sub-flows for each application. To control data 
transmission through an interface, we have to control the sub-flows' priorities over that pair of 
source-destination interfaces. To change the priority of individual sub-flow, we have 
developed new socket API (discussed in \ref{subflow}) where an application can dynamically change 
the priority of an underlying sub-flow.

\begin{figure}
\centering
\includegraphics[scale=0.45]{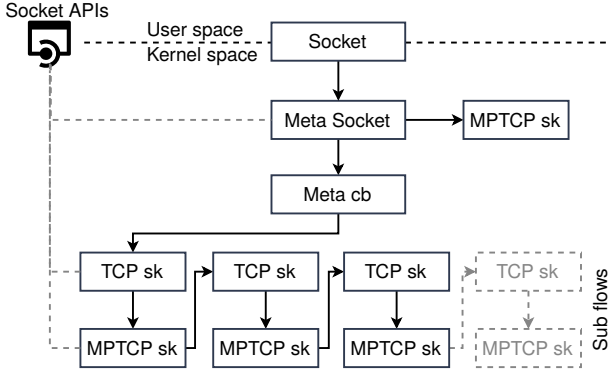}
\caption{MPTCP architecture with socket APIs}
\label{fig:mptcp_socket_API}
\end{figure}

\subsection{Changing sub-flow priority}
\label{subflow}
To change the priority of a sub-flow, we introduce a new socket APIs named 
\texttt{MPTCP\_SET\_SUB\_PRIO}. With this API, an application can make a sub-flow active or backup 
(i.e. high or low priority). However, to use this API, an application need to know the available 
sub-flows in the system. This information can be obtained using socket APIs described in 
\cite{hesmans2016enhanced}. Once an application gets the list of available sub-flows, it can make 
the corresponding sub-flow as backup or active by calling \texttt{setsockopt} with option name 
\texttt{MPTCP\_SET\_SUB\_PRIO} and value as pointer to structure 
\texttt{mptcp\_sub\_prio} from the application as follows:

\begin{lstlisting}
struct mptcp_sub_prio{
	__u8 id;
	__u8 low_prio;
};

struct mptcp_sub_prio flow_prio = {5, 1};
setsockopt(clientSocket, IPPROTO_TCP, \
	MPTCP_SET_SUB_PRIO, \
	&flow_prio, \
	sizeof(flow_prio));
\end{lstlisting}

Here, \texttt{id} is the internal sub-flow id and the \texttt{low\_prio} is the low\_prio flag of 
the sub-flow. To make a sub-flow backup or active, application have to pass \texttt{low\_prio=1} or 
\texttt{low\_prio=0} respectively. On receiving this option,  MPTCP sets the sub-flow 
priority accordingly and also sends this information to the remote host using \texttt{MP\_PRIO} 
header option. When remote host receives \texttt{MP\_PRIO}, it updates sub-flow priority 
accordingly.

Although, \texttt{MPTCP\_SET\_SUB\_PRIO} provides control over a sub-flow priority, there are few 
limitations of this API due to the current implementation of the MPTCP framework. In MPTCP, it is 
not 
possible to remember the priority of any sub-flow between a particular pair of source-destination 
interfaces. 
Hence, if due to some network issues, one of the sub-flows gets destroyed and is replaced by a new 
sub-flow, it will be registered as an active sub-flow irrespective of its earlier priority. The 
application has to call the \texttt{MPTCP\_SET\_SUB\_PRIO} API again to configure the 
new sub-flow. At 
the same time, it should be noted that application is unaware of this phenomenon. Hence, the 
application has to keep track of the sub-flows' status by repeatedly querying the same. This will 
add to a significant overhead at the application layer. In next subsection, we have described a 
possible modification in MPTCP kernel to handle this problem.

\subsection{Remembering sub-flow priority}
\label{remsubflow}
To solve the problem discussed above, we propose to include two lists within the MPTCP
implementation, named \texttt{ActiveInterfaceList} and \texttt{BackupInterfaceList}. With the help 
of these two 
lists, an application can mark a source-destination interface pair as `active' or `backup' 
respectively for a particular application. These are persistent list i.e. once the lists are 
populated, MPTCP follows the list every time it creates a new sub-flow unless the entries in the 
list 
are changed explicitly by the application. The behavior of these two lists is described as follows:

\begin{enumerate}
    \item \texttt{ActiveInterfaceList:} If this exists, then all the sub-flows through the pair of 
interfaces listed here will be active sub-flows and rest of the sub-flows will be backup sub-flows.
    \item \texttt{BackupInterfaceList:} This list contains the pair of source-destination 
interfaces which are supposed to be backup i.e. sub-flows created between these pairs are marked as 
backup while all other sub-flows will be active sub-flows.
    \item Among these two lists, \texttt{ActiveInterfaceList} has higher priority, i.e. if there is 
a common pair exists in both the lists, the entry in the \texttt{ActiveInterfaceList} will get the 
precedence and all the sub-flows created through this pair will be marked as active.
\end{enumerate}

We have also developed socket APIs to manage and maintain these lists. To populate 
\texttt{ActiveInterfaceList} and \texttt{BackupInterfaceList}, an application has to call 
\texttt{setsockopt} with 
option name \texttt{MPTCP\_SUB\_PATH\_ACTIVE\_LIST} and \texttt{MPTCP\_SUB\_PATH\_BACKUP\_LIST} 
respectively. It needs to pass an object of \texttt{struct mptcp\_sub\_path} to 
\texttt{setsockopt} API. Structure of \texttt{struct mptcp\_sub\_path} is as given below:

\begin{lstlisting}
struct mptcp_sub_path{                                                                              
             
    sa_family_t sa_family;
    union{
        struct in_addr sin_addr;
        struct in6_addr sin6_addr;
    };
    union{
        struct in_addr din_addr;
        struct in6_addr din6_addr;
    };
};
\end{lstlisting}

Using these APIs, the lists can be modified any time during the application lifetime. 
However, changing the list does not change the property of any existing sub-flow. Hence, this API 
needs to be called before any sub-flow is created. We have developed a kernel patch for it and the 
same is submitted to the MPTCP-dev mailing list.

\section{Unmanned Aerial Vehicle with MPTCP}
\label{drone}

It is not far into future that Unmanned Aerial Vehicles (UAVs) or drones will be a commonplace like 
cars and aeroplanes. The UAVs require utmost reliability in terms of data communication, 
controlling as well as improved throughput for its high definition live video feed from the camera 
on board. UAVs usually have multiple streams to be communicated to the base station e.g. live 
camera feed, and sensor feed etc. At the same time several control messages need to be delivered to 
the UAV for its operation. For the reliable operation of a UAV, control messages must reach to 
the UAV in a time-bound manner. Any delay in control message may be fatal for the UAV. 
Currently, both these data streams and the control messages are being carried on the same wireless 
link (e.g. same frequency band in WiFi). Hence, it is possible that the live feed from the UAV might 
congest the link and cause significant delay in the delivery of control messages to the UAV 
leading to the failure of the mission or damage to the UAV. 

\begin{figure}[h!]
\centering
\includegraphics[scale=0.7]{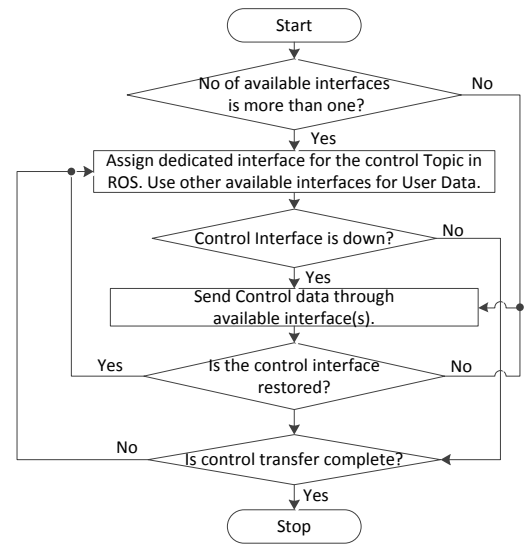}
\caption{Primary Path only Scheduler flow diagram}
\label{fig:sched}
\end{figure}

In this paper, we propose to segregate control data from other user data on separate links 
(through different physical interface) using MPTCP. MPTCP inherently improves the reliability 
of the communication by providing resilience to the link failure. Note that in MPTCP the similar 
effect can be achieved by declaring a network interface as backup using \texttt{ip link} command. 
However, that setting will be system wide and will adversely impact all applications running on that 
system. Hence, we have proposed, \emph{Primary Path only Scheduler} \emph{PPoS}, a new scheduler for 
MPTCP. This ensures that control data is carried through a separate interface and the user data 
does not congest the link allocated for control messages for the particular application 
e.g. UAV in this scenario. At the same time, it retains other inherent properties of MPTCP to 
provide better error resilience to the link failure. \emph{PPoS} achieves the same by marking one or 
more of the MPTCP sub-flows as the ``Primary Path(s)'' ($PP$) for the given application. The 
application continues to use $PP$ only for defined type of data e.g. control data in our case, as 
long as the $PP$ is alive and falls back to the alternative path(s) in case of $PP$ failure. 
However, the transmission is restored to $PP$ as soon as the same is restored 
(Fig.~\ref{fig:sched}). This proposed scheduler is being implemented in the MPTCP Linux kernel for  
testing its performance.

While we have considered UAV as the use case for our proposed scheduler (\emph{PPoS}), it is 
usable for any application which requires to segregate different types of data based upon 
different QoS requirements. 

\vspace{-0.65cm}

\subsection{Socket API for Primary Path only Scheduler (\emph{PPoS})}
We have also developed socket APIs to control the $PP$ selection from the application layer. We 
have introduced a flag named \texttt{primary\_path\_only} in MPTCP kernel to enable \emph{PPoS}. 
Once \emph{PPoS} is enabled, it ensures that all sub-flows other than the one selected are backup 
sub-flows for the particular application. Note that unlike the current MPTCP, these changes affect 
the calling application only. In the event that active sub-flow is not available, \emph{PPoS} 
seamlessly switches to the backup sub-flow(s) for data transmission transparent to the 
application.

\noindent By default \emph{PPoS} is disabled for MPTCP. To enable \emph{PP0S}, one can call 
\texttt{setsockopt} function with option name \texttt{MPTCP\_PRIMARY\_PATH\_ONLY} just after the 
creation of the socket.

\subsection{Integration with Robot Operating System (ROS)}
As discussed in earlier sections, for UAV communication \emph{PPoS} can be very useful. However, 
this requires that these socket APIs are integrated with Robot Operating System (ROS). ROS uses a 
abstracted version of TCP called $ROSTCP$. ROSTCP exposes both python as well as C++ APIs to 
be consumed by other ROS based applications. Integration of MPTCP with ROS is straight forward 
because MPTCP exposes the same socket interfaces to the application as being exposed by TCP. 
However, we have modified $ROSTCP$ suitably to enable/disable \emph{PPoS} from the ROS layer. In 
ROS, if a user wants to use \emph{PPoS} scheduler, it can simply declare an 
environment variable named \texttt{ROS\_MPTCP\_PRIMARY\_PATH\_ONLY}. Our modified $ROSTCP$ 
will enable \emph{PPoS} for this application. Post that the data on $PP$ will not be 
interfered by other sub-flows of MPTCP. At the same time if the $PP$ fails, unlike TCP, the session 
does not get interrupted rather it will move seamlessly to other available sub-flows. Once the 
$PP$ is restored the transmission is restored back to the original sub-flow.

\section{Experiment Results}
\label{exp}

\begin{figure}[h!]
\centering
\includegraphics[scale=0.6]{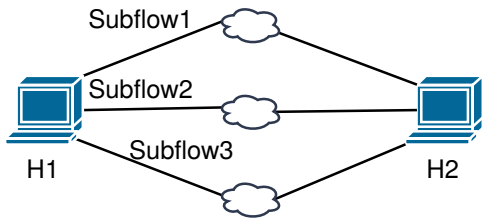}
\caption{Experimental Topology}
\label{fig:topology}
\end{figure}

We have run extensive tests both in the Mininet based environment as well as Raspberry Pi based 
testbed. To test our proposed APIs, we have performed extensive experiments using virtual 
environment created using Mininet\footnote{http://mininet.org/} network emulator. For 
these experiments, we created a simple topology with a pair of source and destination with three 
distinct paths between them (Fig.~\ref{fig:topology}). The bandwidth and end-to-end delay of three 
links are {1Mpbs, 100 ms}. We have assumed the links to be lossless for our experiments.

\vspace{-0.7cm}

\subsection{Changing sub-flow priority}
\label{dprio}
\begin{figure}[h!]
\centering
\includegraphics[scale=0.5]{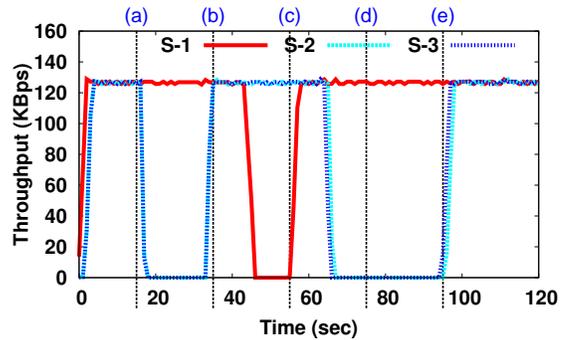}
\caption{State of flows with changing priorities of sub-flows. \textbf{(a)} Mark S-2 and S-3 as
backup, \textbf{(b)} S-1 is disabled, \textbf{(c)} S-1 is enabled, \textbf{(d)} S-2 and S-3 
are disabled, and \textbf{(e)} S-2 and S-3 are enabled.}
\label{fig:plot_priority}
\end{figure}

Firstly, we perform an experiment to study the effect of change of sub-flow priority using our 
developed socket APIs for the topology described in Fig.~\ref{fig:topology}. At the start, all the 
sub-flows (S-1, S-2, S-3) are active and are carrying the data by default. At time 
t=15s (Fig.~\ref{fig:plot_priority}a), we have changed sub-flows S-2 and S-3 as the backup 
sub-flows. As we can notice that after this change, only S-1 is carrying 
the data and both S-2 and S-3 are idle. Again at t=35s (Fig.~\ref{fig:plot_priority}b), we disable 
the active sub-flows (S-1). From this time onwards, backup sub-flows start the data transfer 
transparent to the application. Further after 20 seconds, i.e. at t=55s 
(Fig.~\ref{fig:plot_priority}c), S-1 is restored and MPTCP again switches the data transfer to S-1. 
At t=75s ((Fig.~\ref{fig:plot_priority}d), we disabled all backup sub-flows (S-2 and S-3). 
It is interesting to note that when we re-enabled S-2 and S-3 at t=95s 
((Fig.~\ref{fig:plot_priority}e), new sub-flows are created. However, as described in Section 
\ref{subflow}, these these sub-flows are not able to maintain their earlier state information (i.e. 
backup sub-flows) and all new sub-flows become active and start participating in the data transfer. 

\vspace{-0.8cm}

\subsection{Remembering sub-flow priority}
\begin{figure}
\centering
\includegraphics[scale=0.5]{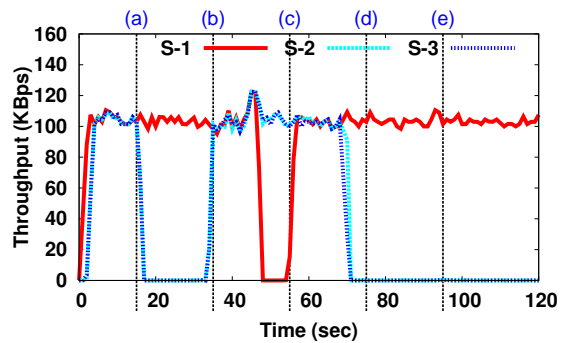}
\caption{State of flows on changing priorities using ActiveInterfaceList and BackupInterfaceList.}
\label{fig:plot_with_list}
\end{figure}

From the experiment in \ref{dprio}, we notice that in the event of disconnection sub-flow(s) are not 
able to remember their state. So, we use \texttt{ActiveInterfaceList} and 
\texttt{BackupInterfaceList} (Section \ref{remsubflow}) to mark the corresponding pair of 
source-destination interfaces as active and backup ones respectively. We perform the same experiment 
as the previous one with the change that we added S-2 and S-3 to the \texttt{BackupInterfaceList}. 
Fig.~\ref{fig:plot_with_list} shows the result for our experiment. We can notice here that after 
re-enabling S-2 and S-3 (Fig.~\ref{fig:plot_with_list}e), new sub-flows do not become active 
again i.e. they remember their earlier state information.

\vspace{-0.3cm}

\subsection{Using \emph{PPoS} Scheduler}
We have conducted another set of experiments with ROS based system using Raspberry Pi boards. We 
have used the same topology as shown in Fig.~\ref{fig:topology}. The sub-flow (S-1) is being chosen 
to carry the control data and other sub-flows are carrying the user data. Here, for the ease of 
representation and clarity, we are only showing the results where the data is being carried on S-1 
only and there is no data on other sub-flows. However, the behavior of the proposed scheduler 
remains the same even if the data is being carried on all sub-flows.

\begin{figure}[!h]
	\captionsetup[subfigure]{}
	\begin{center}
		\subfloat[\label{fig:plot_default} With default scheduler]{
			\includegraphics[width=0.44\linewidth]{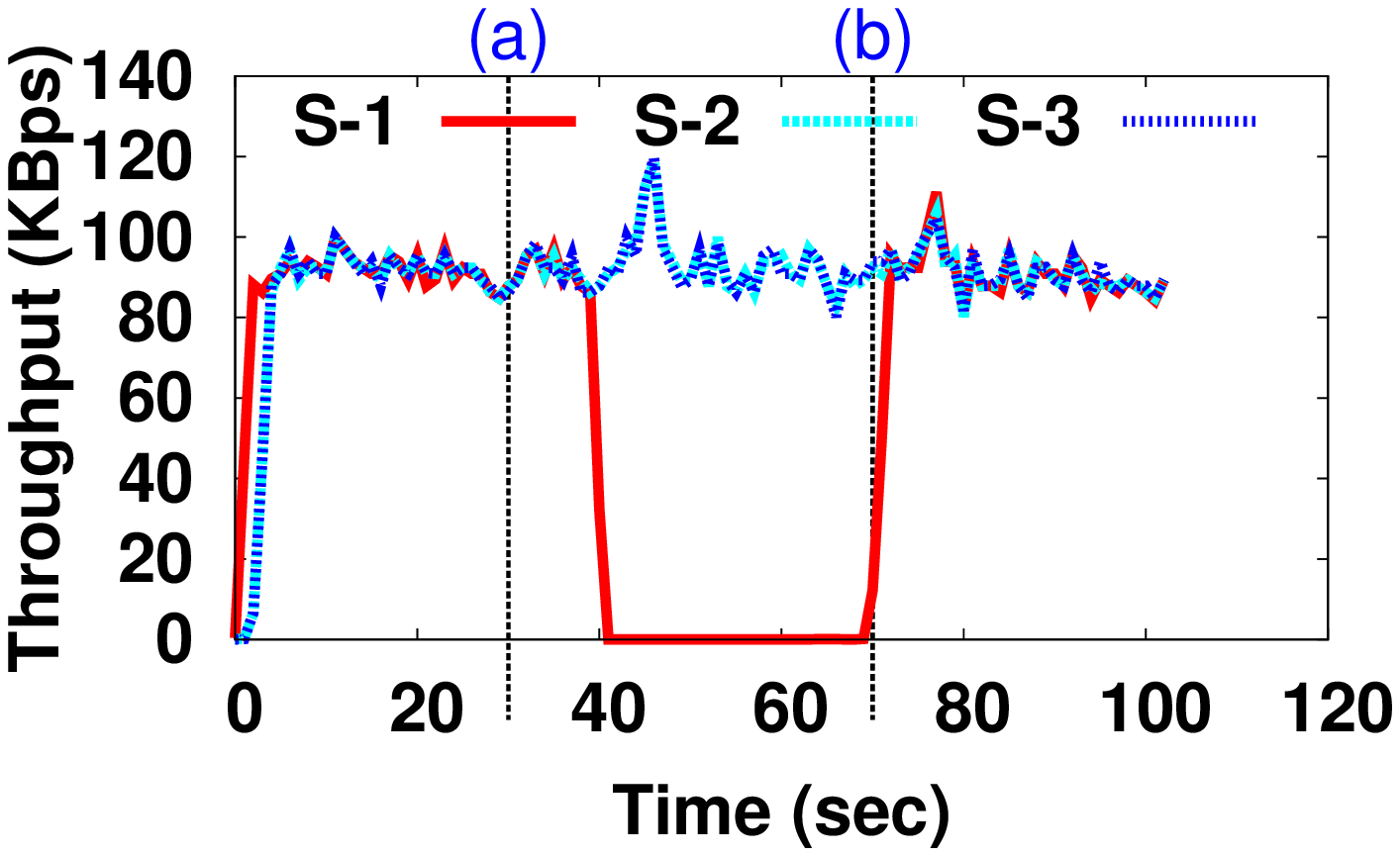}
		}
		\subfloat[\label{fig:plot_primary_only} With \emph{PPoS} scheduler]{
			\includegraphics[width=0.44\linewidth]{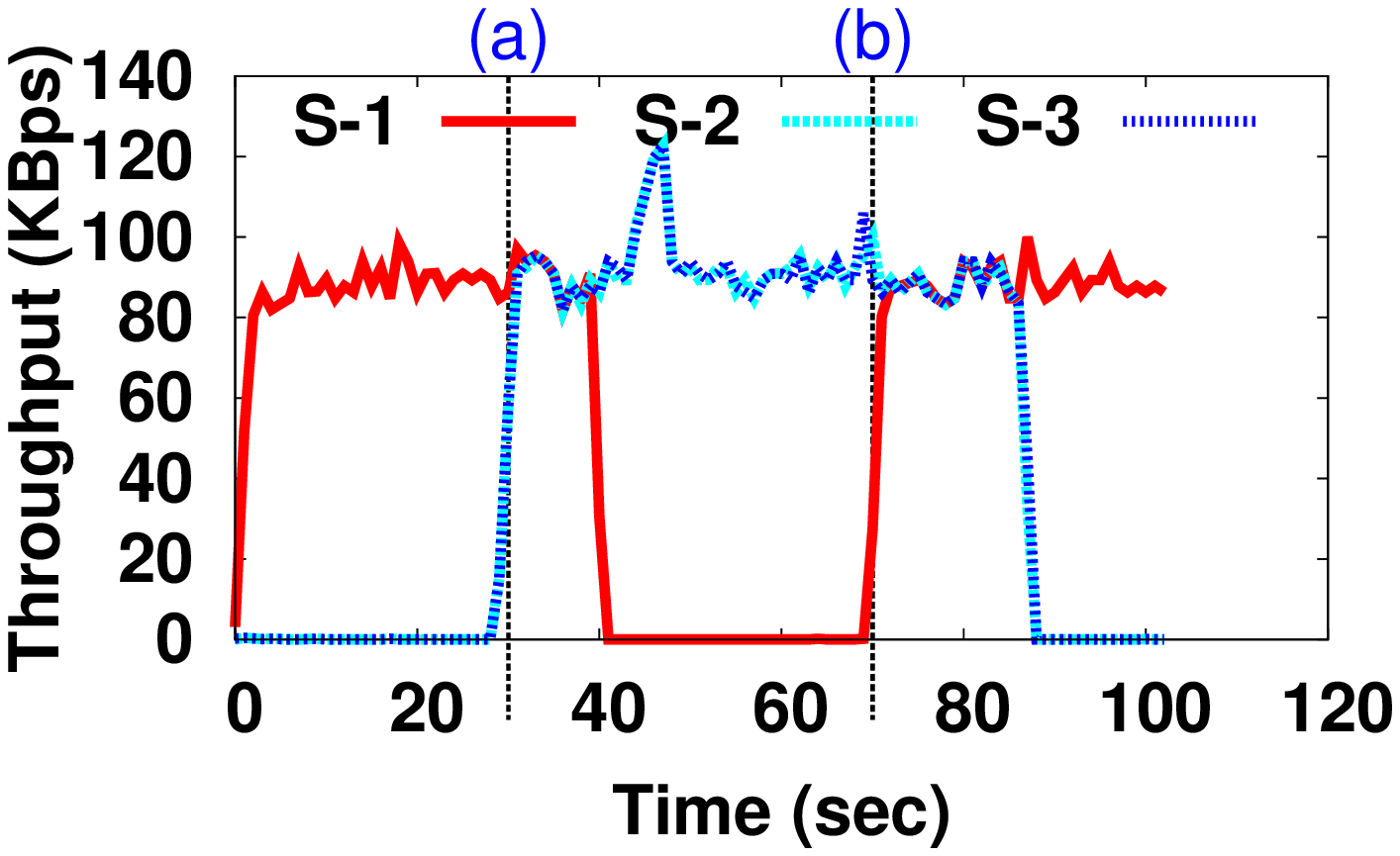}
		}
		\caption{\label{fig:compare_path} Comparing MPTCP default scheduler and \emph{PPoS} 
scheduler. At (a) S-1 is disabled and at (b) S-1 is restored.}
	\end{center}
\end{figure}

In Fig.~\ref{fig:compare_path}, we have compared the default behaviour of MPTCP 
(Fig.~\ref{fig:compare_path}a) with \emph{PPoS} (Fig.~\ref{fig:compare_path}b). In these 
experiments, we are transmitting data between hosts H1 and H2 for 100 seconds. For 
both experiments, we dropped S-1 (sub-flow for carrying control data) at the time t=30s and 
re-enabled the same at t=70s. As we can note that with default settings, MPTCP continues to send 
the data on all sub-flows all the time. With \texttt{MPTCP\_PRIMARY\_PATH\_ONLY} option 
being enabled at ROS, MPTCP sends data only on the selected $primary path$ and does not send any 
data on other sub-flows. However, as S-1 goes down, it starts using the other sub-flows for 
transmission. As soon as S-1 restores, the transmission is also restored to originally selected 
primary path (S-1). This feature of \emph{PPoS} makes it an interesting choice 
for scenarios and applications such as UAVs where control messages should be provided utmost 
reliability and delay in delivery of control messages may prove fatal.

\renewcommand\thesubfigure{\roman{subfigure}}

\vspace{-0.8cm}

\section{Conclusion}
\label{concl}

MultiPath TCP is a way to utilize multiple network simultaneously and can support handover between 
different networks seamlessly. After being implemented for the Linux kernel, it has been ported to 
various devices and architectures. Although, it is a drop-in replacement of the standard TCP, it 
lacks socket APIs to control and modify its functionalities from the application layer. The 
control through kernel parameter is a system wide change and many a times is not desirable. We have 
designed and developed several socket APIs to control sub-flows' priorities from the application 
layer. We have presented Primary Path only Scheduler (\emph{PPoS}), a new  scheduler for the 
applications and devices (e.g. UAVs) which require to segregate the data on multiple interfaces for 
various reasons such as different QoS requirements and reliability etc. We have also developed the 
socket APIs for this scheduler and implemented them in the MPTCP Linux kernel. Using these socket 
APIs, we can selectively enable the scheduler and control its behaviour only for specific 
applications rather than doing it system wide. 

\bibliographystyle{IEEEtran}
\bibliography{ref/api}

\end{document}